\documentclass[pre,twocolumn,superscriptaddress]{revtex4}

\usepackage[dvips]{graphicx}

\begin{document}

\preprint{}
\title[]{Nonmonotonic reversible branch in four model granular beds subjected to vertical vibration.}
\author{Luis A. Pugnaloni\footnote{Email: luis@iflysib.unlp.edu.ar}}
\affiliation{Instituto de F\'{\i}sica de L\'{\i}quidos y Sistemas Biol\'{o}gicos (CONICET, UNLP), cc. 565, 1900 La Plata, Argentina.}
\author{Mart\'{\i}n Mizrahi}
\affiliation{Departamento de F\'{\i}sica, Facultad de Ciencias Exactas, Universidad Nacional de La Plata, cc. 67, 1900 La Plata, Argentina}
\author{Carlos M. Carlevaro}
\affiliation{Instituto de F\'{\i}sica de L\'{\i}quidos y Sistemas Biol\'{o}gicos (CONICET, UNLP), cc. 565, 1900 La Plata, Argentina.}
\affiliation{Universidad Tecnol\'{o}gica Nacional, FRBA, UDB F\'{\i}sica, Mozart 2300, C1407IVT, Buenos Aires, Argentina}
\author{Fernando Vericat}
\affiliation{Instituto de F\'{\i}sica de L\'{\i}quidos y Sistemas Biol\'{o}gicos (CONICET, UNLP), cc. 565, 1900 La Plata, Argentina.}
\affiliation{Grupo de Aplicaciones Matem\'{a}ticas y Estad\'{\i}sticas de la Facultad de Ingenier\'{\i}a (GAMEFI), UNLP, La Plata, Argentina.}

\begin{abstract}
We present results from four independent models of a granular assembly subjected to tapping. We find that the steady state packing fraction as a function of the tapping intensity is nonmonotonic. In particular, for high tapping intensities, we observe an increase of the packing fraction with tapping strength. This finding challenges the current understanding of compaction of granular media since the steady state packing fraction is believed to decrease monotonically with increasing tapping intensity. We propose an explanation of our new results based on the properties of the arches formed by the particles.
\end{abstract}

\volumeyear{}
\volumenumber{}
\issuenumber{}
\eid{}
\date{September 2008}
\startpage{1}
\endpage{}
\maketitle

\section{Introduction}
The study of compaction of granular matter under vertical tapping is a subject of much debate and consideration \cite{Richard1}. Most studies nowadays focus on the dynamics of compaction --the evolution of structural properties as a function of the number of taps applied to the sample. In spite of being of chief importance, much less work is done on the steady state regime achieved by the sample after a very large number of taps. One of the reasons for this is the fact that at very low tapping intensities --when the entire granular bed does not detach from the containers bottom upon tapping-- the relaxation dynamics of these systems is extremely slow, which makes the steady state very hard to reach. In a pioneering work, Nowak et al. \cite{Nowak1} showed that the steady state can be achieve by means of a suitable annealing. Very recently Ribi\`{e}re et al. \cite{Ribiere1} argued that the steady state is indeed obtainable, reproducible and may constitute a true thermodynamic state for granular systems. These experiments show that the packing fraction $\phi$ in the steady state is a monotonic decreasing function of the tapping intensity. However, to our knowledge, all these studies have explored a limited range of tapping intensities. High tapping intensities --reduced acceleration $\gg 1$-- are difficult to handle in the laboratory due to the height reached by the surface particles during a strong tap. In this paper we present simulations coming from four different models that explore the high intensity tapping regime of the steady state. The models describe the packing of grains subjected to tapping to different degrees of complexity. We find that the packing fraction $\phi$ is nonmonotonic. The fact that all four models show the same general trend supports the claim that the phenomenon is robust and should be seen in most experimental samples. This constitutes a challenge to many simpler models that aim at explaining the underlying mechanisms of granular compaction. We present a first attempt of explanation of the new results based on the formation of arches in the packing.

\section{Models}
The models used in this work have been introduced previously in the literature. Here we summarize the most important features and technical details that are relevant to this study. A more detailed description of each model can be found in the original papers.

\subsection{Hybrid Monte Carlo + ballistic deposition} 
This model has been introduced over a decade ago \cite{Barker1}. A packing of spheres is compressed in a uniaxial external field using a low-temperature Monte Carlo process. Then, the spheres are stabilized using a steepest decent drop-and-roll dynamics to find a local minimum of the potential energy. During this second phase of the restructuring, the spheres, although moved in sequence, are able to roll in contact with spheres that are in either stable or unstable positions. In this way, mutual stabilization may arise. The final configuration has a well defined network of contacts and each sphere has a uniquely defined three point stability.

After deposition is completed, a homogeneous expansion that simulate the introduction of free volume caused by tapping is induced by rescaling the vertical coordinates of the spheres by a factor $A>1$. Then, a new MC + ballistic deposition takes the system to the next stable configuration.

\subsection{Pseudodynamics} 
In this model the entire deposition is conducted using a ``simultaneous'' ballistic deposition of hard disks \cite{Manna1,Manna2,Pugnaloni1}. In practice, disks fall or roll one at a time, but only by short distances so as to mimic a more realistic dynamics than in model A.

The deposition algorithm consists in picking up a disk in the system and let it perform a free fall of length $\delta$ if the disk has no supporting contacts, or a roll of arclength $\delta$ over its supporting disk if the disk has one single supporting contact. Disks with two supporting contacts are considered stable and left in their positions. If in the course of a fall of length $\delta$ a disk collides with another disk or the base of the container, the falling disk is put just in contact and this contact is defined as its first supporting contact. Analogously, if in the course of a roll of length $\delta$ a disk collides with another disk or a wall, the rolling disk is put just in contact. If the first supporting contact and the second contact are such that the disk is in a stable position, the second contact is defined as the second supporting contact; otherwise, the lowest of the two contacting particle is taken as the first supporting contact of the rolling disk and the second supporting contact is left undefined. If, during a roll, a particle reaches a lower position than the supporting particle over which it is rolling, its first supporting contact is left undefined. A moving disk can change the stability state of other disks supported by it; therefore, this information is updated after each move. The deposition is over once each particle in the system has both supporting contacts defined.

As in model A, tapping is simulated by scaling the vertical coordinates of the particles by a factor $A>1$. Since this model lacks a MC phase, the random rearrangements produced by tapping are simulated by introducing a few random displacements after expansion and before the next pseudodynamic deposition.

\subsection{Monte Carlo}
This is a very simple model that only considers the excluded volume of the particles \cite{Philippe1}. A MC deposition of the particles is used in the same way as in model A. However, there is not a ballistic deposition to enssure that particles come to contact. At each MC step each particle is given a chance to make a random move. Moves are only accepted if they do not lead to an overlap. Each random displacement is generated by choosing a uniform random vector in the interval $x\in[-0.01d,0.01d]$, $y\in[-0.01d,0.01d]$ and $z\in[-0.01d,0.0]$. Here, $d$ is the diameter of the particles.

In Ref. \cite{Philippe1} each deposition is terminated once the average height of the particles does not evolve. Strong taps, like the ones we study in this work, have the drawback that some particles at the top may have not join the rest of the pack by the time the average height of the pack has levelled off due to the bulk of the system being very dense. We then use a termination criterion that requires that the acceptance ratio of each individual particle falls below a threshold. We simulate hard spheres and hard disks using this very simple model.
 
\subsection{Granular Dynamics} 
This is the most realistic model we present \cite{Arevalo1}. We use a soft-particle two-dimensional molecular dynamics (MD). Particle--particle interactions are controlled by the particle--particle overlap $\xi =d-\left\vert \mathbf{r}_{ij}\right\vert $ and the velocities $\dot{\mathbf{r}}_{ij}$, $\omega _{i}$ and $\omega _{j}$. Here, $\mathbf{r}
_{ij}$ represents the center-to-center vector between particles $i$ and $j$, $d$ is the particle diameter and $\omega $ is the particle angular velocity. These forces are introduced in the Newton's translational and rotational equations of motion and then numerically integrated by standard methods \cite{Schafer1}. 

The contact interactions involve a normal force $F_{\text{n}}$ and a tangential force $F_{\text{t}}$. We use a normal force which involves a linear (Hookean) interaction between particles.

\begin{equation}
F_{\text{n}}=k_{\text{n}}\xi -\gamma _{\text{n}}v_{i,j}^{\text{n}}
\label{normal}
\end{equation}

\begin{equation}
F_{\text{t}}=-\min \left( \mu |F_{\text{n}}|,|F_{\text{s}}|\right) \cdot 
\text{sign}\left( \zeta \right)  \label{tangent}
\end{equation}%
where

\begin{equation}
F_{\text{s}}=-k_{\text{s}}\zeta -\gamma _{\text{s}}v_{i,j}^{\text{t}}
\label{contact}
\end{equation}

\begin{equation}
\zeta \left( t\right) =\int_{t_{0}}^{t}v_{i,j}^{\text{t}}\left( t^{\prime
}\right) dt^{\prime }  \label{static}
\end{equation}

\begin{equation}
v_{i,j}^{\text{t}}=\dot{\mathbf{r}}_{ij}\cdot \mathbf{s}+\frac{1}{2}d\left(
\omega _{i}+\omega _{j}\right)  \label{vtan}
\end{equation}

The first term in Eq.~(\ref{normal}) corresponds to a restoring force proportional to the superposition $\xi $ of the interacting disks and the stiffness constant $k_{n}$. The second term accounts for the dissipation of energy during the contact and is proportional to the normal component $v_{i,j}^{\text{n}}$ of the relative velocity $\dot{\mathbf{r}}_{ij}$ of the
disks.

Equation~(\ref{tangent}) provides the magnitude of the force in the tangential direction. It implements the Coulomb's criterion with an effective friction following a rule that selects between static or dynamic friction. Notice that Eq. (\ref{tangent}) implies that the maximum static friction force $|F_{\text{s}}|$ used corresponds to $\mu |F_{\text{n}}|$, which effectively sets $\mu_{\text{dynamic}}=\mu_{\text{static}}=\mu$. The static friction force $F_{\text{s}}$ [see Eq. (\ref{contact})] has an
elastic term proportional to the relative shear displacement $\zeta $ and a dissipative term proportional to the tangential component $v_{i,j}^{\text{t}} $ of the relative velocity. In Eq. (\ref{vtan}), $\mathbf{s}$ is a unit vector normal to $\mathbf{r}_{ij}$. The elastic and dissipative contributions are characterized by $k_{\text{s}}$ and $\gamma _{\text{s}}$
respectively. The shear displacement $\zeta $ is calculated through Eq. (\ref{static}) by integrating $v_{i,j}^{\text{t}}$ from the beginning of the contact (i.e., $t=t_{0}$). The tangential interaction behaves like a damped spring which is formed whenever two grains come into contact and is removed when the contact finishes \cite{wolf}.

Tapping is simulated by applying an external vertical motion to the container of the form of half sine wave [$z_0 sin(\omega t)$]. Intensity is controlled by the amplitude of the excitation so as to obtain different reduced accelerations $\Gamma=z_0 \omega^2 / g$ (with $g$ the acceleration of gravity). The interaction of the particles with the flat surfaces of the container is calculated as the interaction with a disk of infinite radius. The particular set of parameters used for the simulation is: $\mu = 0.5$, $k_n = 10^5 (mg/d)$, $\gamma_n = 300 (m\sqrt{g/d})$, $k_s= \frac{2}{7}k_n$ and $\gamma_s = 200 (m\sqrt{g/d})$. The integration time step is set to $\delta = 10^{-4} \sqrt{d/g}$, and $m$ is the mass of the particles. 

\section{The relation between the models}
\subsection{Tapping protocol}
\begin{figure}
\includegraphics[width=0.45\textwidth]{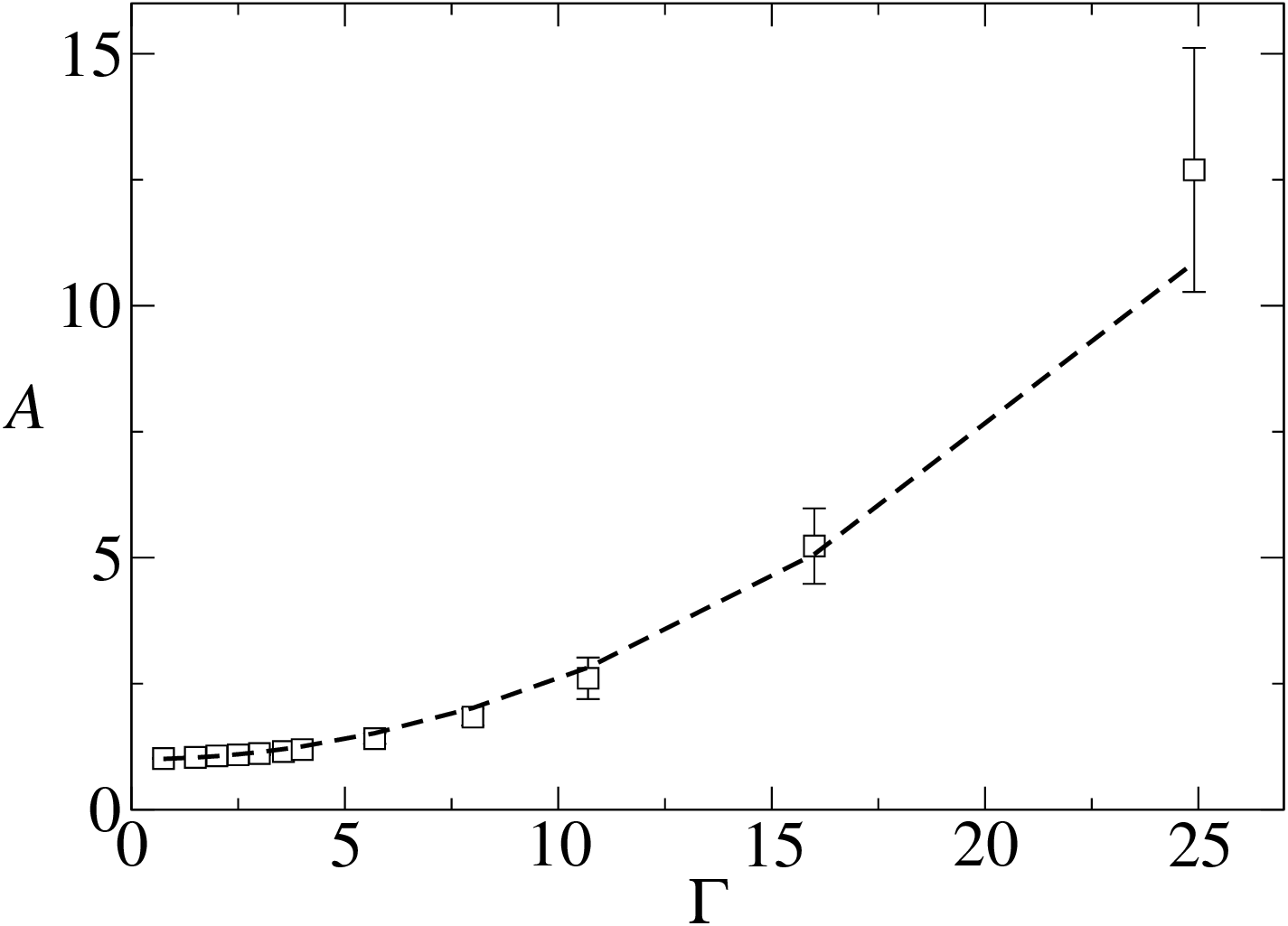}
\caption{\label{fig1} Expansion $A$ as a function of the reduced acceleration $\Gamma$ observed in the granular dynamics model D. The fit corresponds to $A=0.0159 \Gamma^2+1$.}
\end{figure}

In models A to C the tapping intensity is parametrized with the expansion factor $A$ instead of a realistic reduced acceleration. Philippe and Bideau \cite{Philippe1} have suggested that this intensity should be related with actual reduced acceleration as $\Gamma = \alpha (A-1)^{1/2}$. We have estimated the expansion factor $A$ in the granular dynamics model D taken the uplift in the position of the center of mass (CM) of the packing during a tap. The maximum position of the CM during the flight of particles after a tap is divided by the position of the CM at rest before the tap is applied. The actual expansion is not homogeneous since, during a tap,  particles at the bottom of the bed depart from each other much less than particles at the top. We use the CM as a simple estimator of the average expansion. In FIG. \ref{fig1} we plot the estimated $A$ as a function of $\Gamma$. The values of $A$ are averages over about 100 taps in the stationary regime. The error vars correspond to the standard deviation. It is clear that the functional dependence suggested by Philippe and Bideau is indeed valid. In the rest of the graphs we use $\varepsilon \equiv (A-1)^{1/2}$ as a measure of the tapping intensity. For model D the values of $\varepsilon$ are obtained from the estimated expansion using the uplift of the CM of the bed.

\subsection{Particle--particle interactions}
Models A to D have been used before to study the compaction of granular samples under tapping and have been shown to be adequate to describe many of the features seen in experiments. However, all four models describe particle-particle interactions to different levels of complexity and we are interested in probing how much detail is necessary to model general features of granular compaction. For this reason we explore the response of all four models to high intensity taping.

The simplest model considered (model C) takes into account solely the excluded volume of the particles in a rather crude fashion. At the end of each deposition, particles are left very close to each other. However, there are no contacts at all in the system. As a consequence, it is not possible to define what are the grains that support a given particle. Moreover, the dynamics of the deposition in rather artificial since real collisions are not simulated.

Model A represents an important improvement to model C, besides being introduced early in the literature. In this case the dynamics of deposition is still a MC type process. However, the deposition is finished off with a ballistic phase that ensures that all particles end up with all the contacts required to make particles stable in their positions. This type of ballistic deposition considers no bounces between the particles and the rolling is always made without sliding. In practice this mimics hard particles with coefficient of restitution zero. Besides, these particles are effectively considered to be of zero mass since they never have a true velocity defined and then momentum conservation is not taken into account in the collisions.

The pseudodynamics of model B avoids to some extent the artificial MC compaction used in models A and C. Nevertheless, this type of dynamics is still rather simplistic since it follows the same assumptions than the ballistic phase of model A, i.e. no bounces between the particles and rolling without sliding. The main improvement with respect to model A is that particles are not moved one at a time until they reach their stable positions, but simultaneously. In all other respects the particle-particle interactions remain the same.

Finally, model D tries to consider all aspects involved in the contact dynamics of soft particles. Although it is necessary to explore a vast range of values for the coefficients used in the force laws in order to asses to what extent the different aspects (restitution coefficient, static friction, dynamic friction, elastic repulsion, etc.) affect the results discussed below, we present results for a single set of parameter. Our aim is to show that the features displayed by the much simpler models A, B and C are still present in this realistic model, and that they are not artifacts of the simplistic approaches.

\section{Results and discussion}
Table \ref{table1} summarizes the parameters used for the simulations in each model. It is important to remark here that the steady state is obtained after a relatively small number of taps in all our simulations since we do not explore the low intensity tap regime \cite{Ribiere1}. Although we present results from a single set of experiments for each curve, we have repeated the simulations up to five times with different initial conditions. Deviations between independent experiments are within the size of the symbols. The same results are found when an annealing tapping is used to obtain the reversible branch of the $\phi$--$\varepsilon$ curves before analyzing the data at each particular value of $\varepsilon$. Simulations with containers of different sizes showed the same trends in all models although the actual values of $\phi$ vary. The same is true when results from systems with periodic boundaries are compared with systems with hard walls. We have also carried out a few simulations using model B where a quadratic, rather than homogeneous, expansion is used for the taps. This type of expansion seems to be more realistic compared with the granular dynamic simulation in the sense that, during a tap, particles at the top of the pile tend to separate from each other much more than particles at the bottom of the container. We have seen that this has very little effect in the results discussed below.

\begin{table}
\caption{\label{table1} Setup of the simulations for the four models. Averages indicate the number of configurations generated and the frequency at which they were sampled for analysis in parenthesis. Equilibration is the number of taps that were necessary to reach the steady state. The unit length is the particle diameter} 
\begin{tabular}{cccccc}
\hline
Model & No. part. & box & boundaries & Equilibr. & Averages \\
\hline
A spheres & 1000 & $7\times 7$ & periodic & 100 & 500(10) \\
\hline
B disks & 2000 & $13.4$ & hard walls & 100 & 1000(10) \\
\hline
C spheres & 864 & $7\times 7$ & periodic & 200 & 700(1) \\
C disks & 1210 & $12.39$ & hard walls & 100 & 900(1) \\
\hline
D disks & 512 & $12.39$ & soft walls & 100 & 400(1) \\
\hline
\end{tabular} 

\end{table}

\begin{figure}
\includegraphics[width=0.45\textwidth]{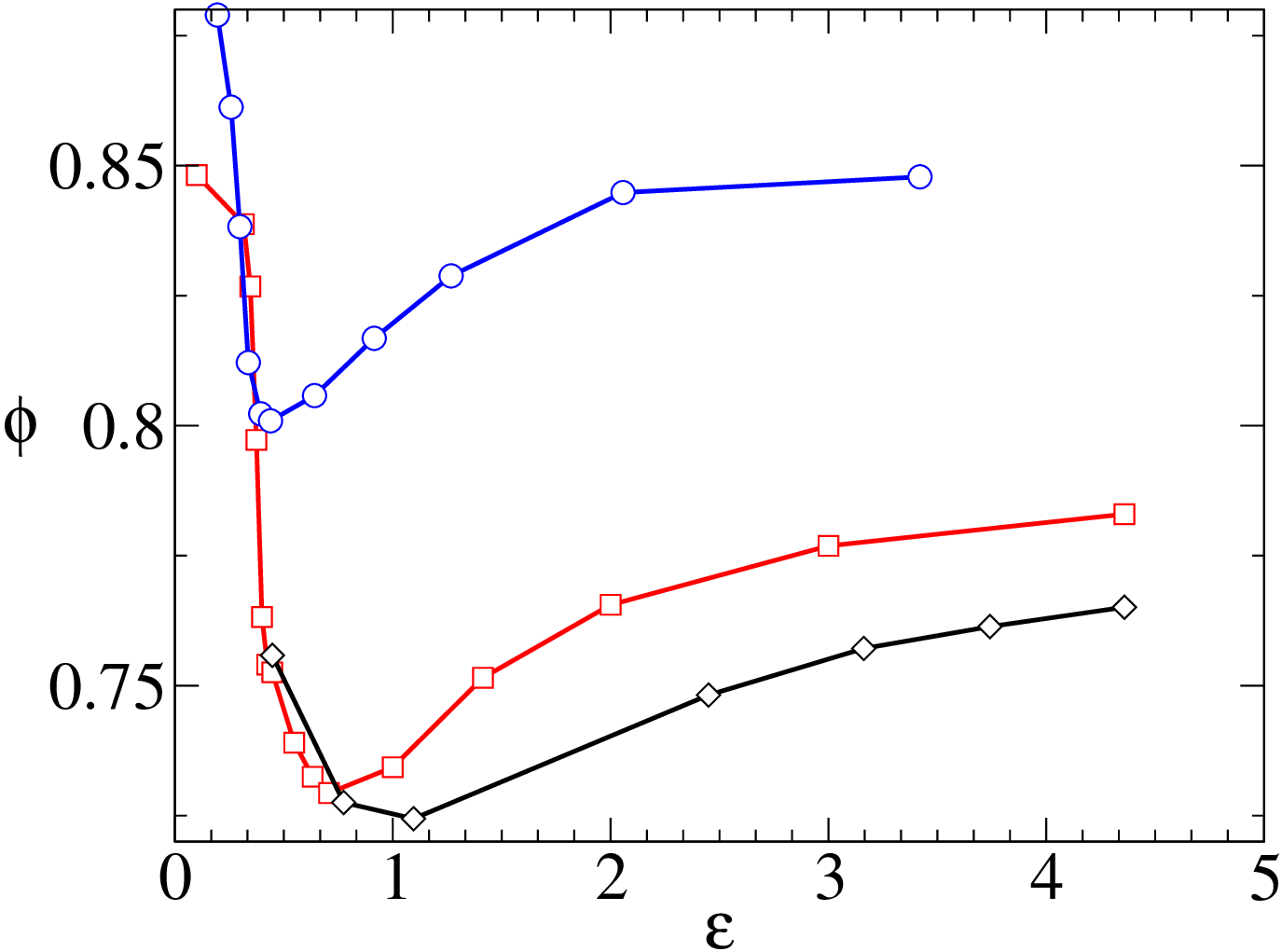}
\caption{\label{fig2} (Color on-line) Steady state packing fraction $\phi$ of disks as a function of tapping intensity $\varepsilon$ for models B (squares), C (diamonds) and D (circles).}
\end{figure}

In FIG. \ref{fig2} we show the packing fraction of disks as a function of the tapping intensity as measured by the parameter $\varepsilon$ for models B, C and D. As we can see, $\phi$ presents a sharp decrease at relatively low values of $\varepsilon$. After that, a further increase in tapping intensity leads to a rise in packing fraction that eventually levels off.

Besides the qualitative agreement, there is a clear gap between the values of $\phi$ from models B and C, and those from model D. Model B is intended to simulate particles with a zero restitution coefficient. Model D, for the particular interaction parameters used, have a coefficient of normal restitution of $e_n=0.058$ \cite{Arevalo1}. In practice, particles in model D can bounce after contacting deposited particles and make these particles to spring back from the bed. This induces rearrangements in the packing that lead to more compact and ordered structures. The range of packing fractions obtained in model D agree rather well with the values found in event-driven simulations of tapping \cite{Arsenovic1}. In model B, on the other hand, particles touch and roll nicely on deposited particles inducing no restructuring at all by this mechanism---which resembles the deposition of particles immersed in a very viscous fluid. This makes structures built with model B more open. A realistic granular dynamics of disks flowing down an incline \cite{Silbert1} have yielded values of packing fractions in the same range as model A. Model C, is a rather simplistic representation of compaction since particles never touch; however, the packing fraction obtained is rather similar to that predicted by model B. We suggest that deposition by MC techniques are then a good model for grains with zero restitution coefficient.

We expect that the plateau at high values of $\varepsilon$ should coincide with the limiting case of sequential deposition of disks. If $\varepsilon$ is very large, particles get well separated from each other during a tap, which makes grains to deposit almost independently. Such limit has been extensively investigated in the past for model B with periodic boundaries in the horizontal direction \cite{Barker2}. The reported packing fraction is $\phi \approx 0.82$. This value is somewhat higher than the packing fraction we have obtained for model B at the larger $\varepsilon$ studied ($\phi \approx 0.785$). We have to bear in mind, however, that the impenetrable walls used in our simulation induce a marked reduction in the packing fraction. We have measured the packing fraction in a rectangle spanning the walls in the horizontal directions and of height equal to 50\% of the bed height centered in the middel of the pile. If instead we measure the packing fraction inside a rectangle of half the container width centered in the simulation box, the maximum value obtained at large $\varepsilon$ becomes $\phi \approx 0.815$, which compares very well with the result from REF. \cite{Barker2}. Sequential depositions carried out with realistic models like model D have not yet been undertaken to our knowledge. These type of simulations are very time consuming since one needs to wait for the entire bed to relax after depositing a particle before the next is released. 

\begin{figure}
\includegraphics[width=0.45\textwidth]{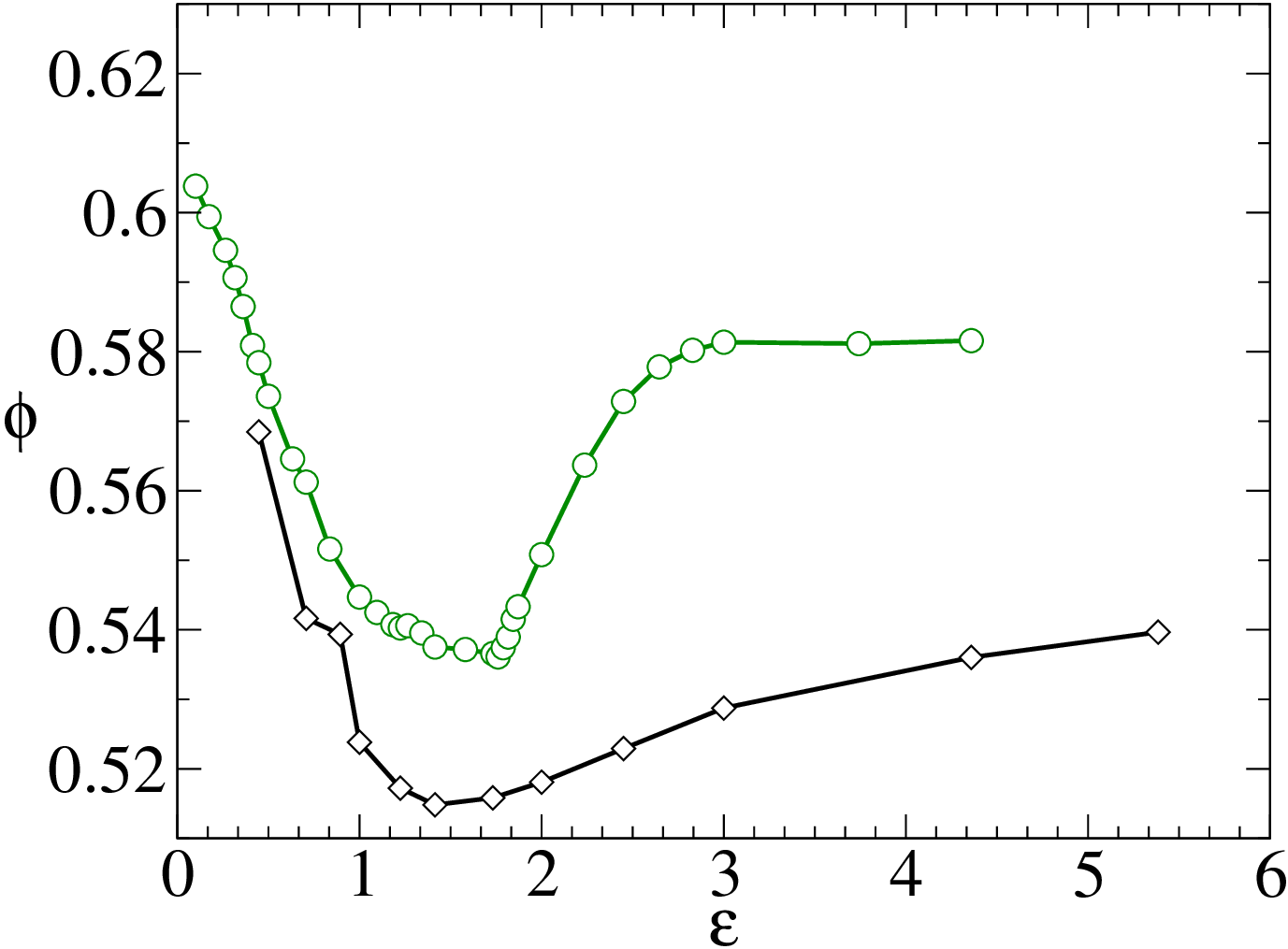}
\caption{\label{fig3} (Color on-line) Steady state packing fraction $\phi$ of spheres as a function of tapping intensity $\varepsilon$ for models A (circles) and C (diamonds).}
\end{figure}

The packing fraction for models A and C applied to spheres is presented in FIG. \ref{fig3}. Once again, the increase and saturation in the packing fraction is clearly observed at large tapping intensities. Model C shows lower densities as compared with model A. These two models differ in that the hybrid MC + ballistic deposition ensure that all particles reach stable positions in contact with others whereas the MC scheme is terminated when acceptance of moves is low; leaving particles ``in the air''. The position of the minimum in the curves is roughly the same in both models. A realistic granular dynamics of spheres allowed to settle under gravity from a dilute configuration lead to $\phi \approx 0.6$ \cite{Silbert2} which is about the maximum packing fraction attained in model A.

Here again, the limiting case of sequential deposition should be attained at very strong tapping intensities. For model A, such sequential deposition has been reported to yield $\phi \approx 0.58$ \cite{Mehta2}. This is indeed the limit reached for our results at $\varepsilon > 3$.

The increase of the steady state packing fraction with tapping intensity at high $\varepsilon$ seems to be a rather fundamental feature since it is present in all models studied irrespective of the details of the simulated tapping and deposition protocols. 

The initial decrease in packing fraction at low $\varepsilon$ is very well known and seemingly understood. As volume injection through tapping is enhanced, it is expected that larger voids may get trapped during deposition which leads to lower packing fractions. However, this line of thinking is unable to explain the appearance of a minimum at large tapping intensities and the increase and saturation of the packing fraction.

\begin{figure}
\includegraphics[width=0.45\textwidth]{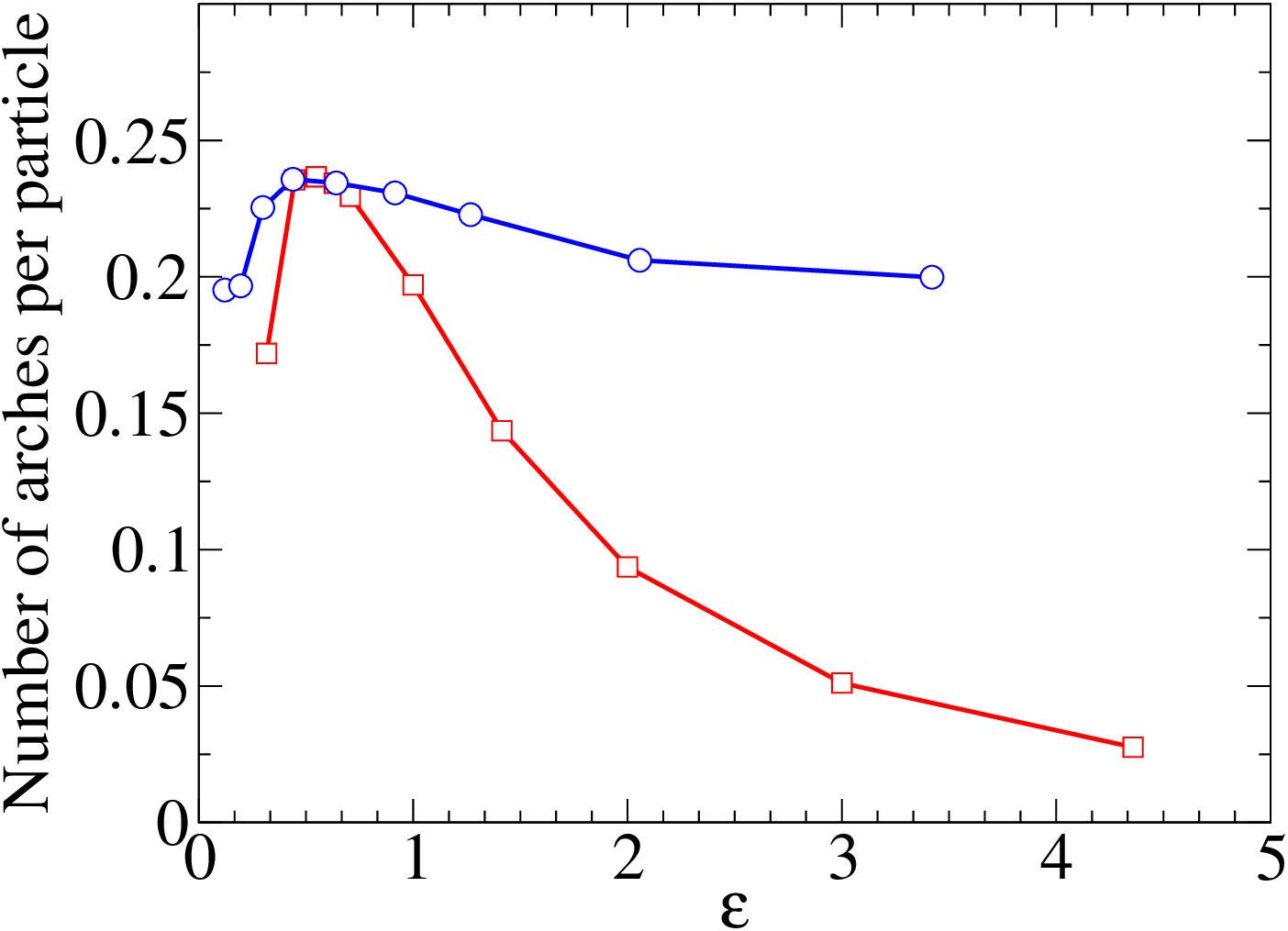}
\caption{\label{fig4} (Color on-line) Steady state number of arches for disks as a function of the tapping intensity $\varepsilon$ for models B (squares) and D (circles).}
\end{figure}

\begin{figure}
\includegraphics[width=0.45\textwidth]{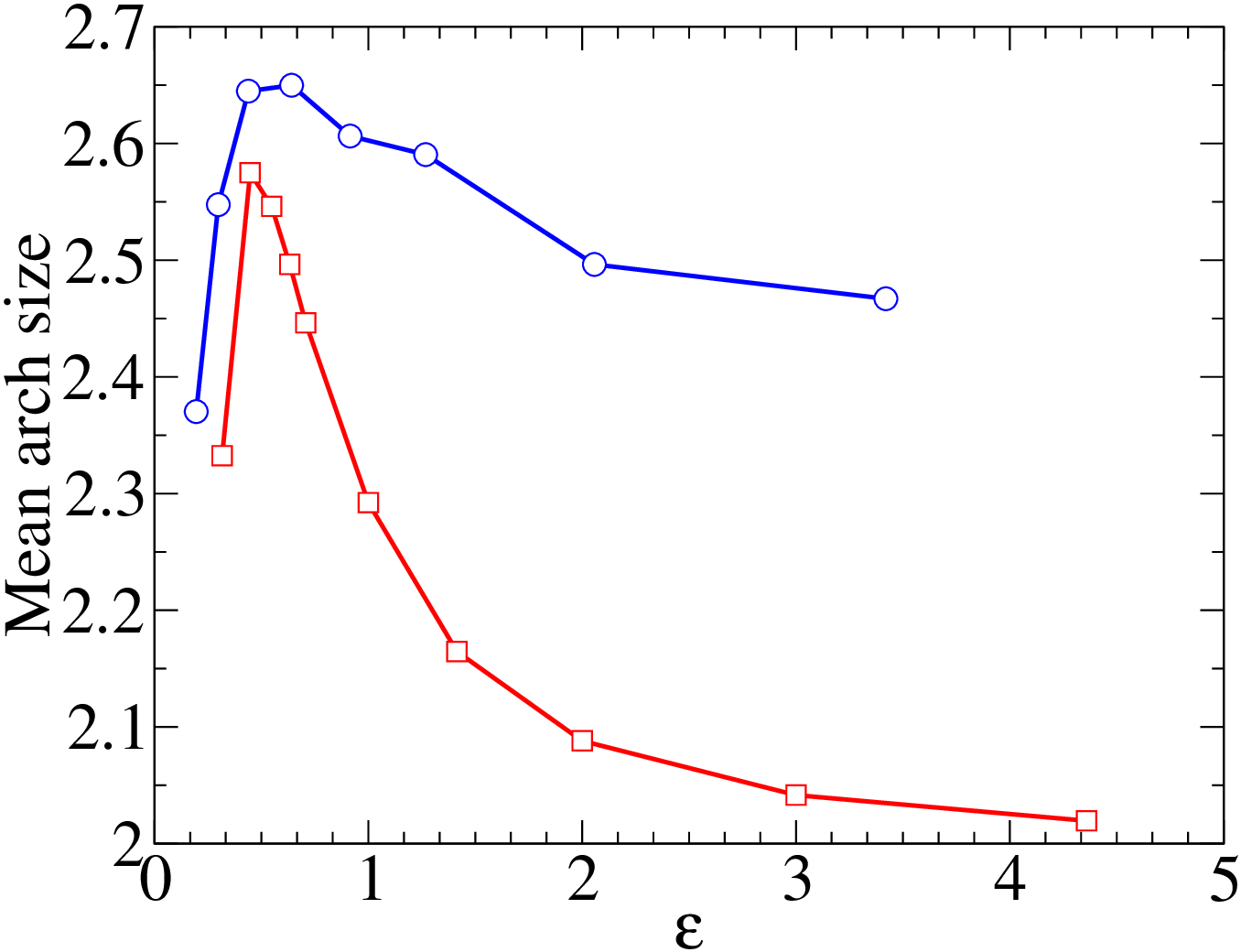}
\caption{\label{fig5} (Color on-line) Steady state mean arch size for disks as a function of the tapping intensity $\varepsilon$ for models B (squares) and D (circles).}
\end{figure}

\begin{figure}
\includegraphics[width=0.45\textwidth]{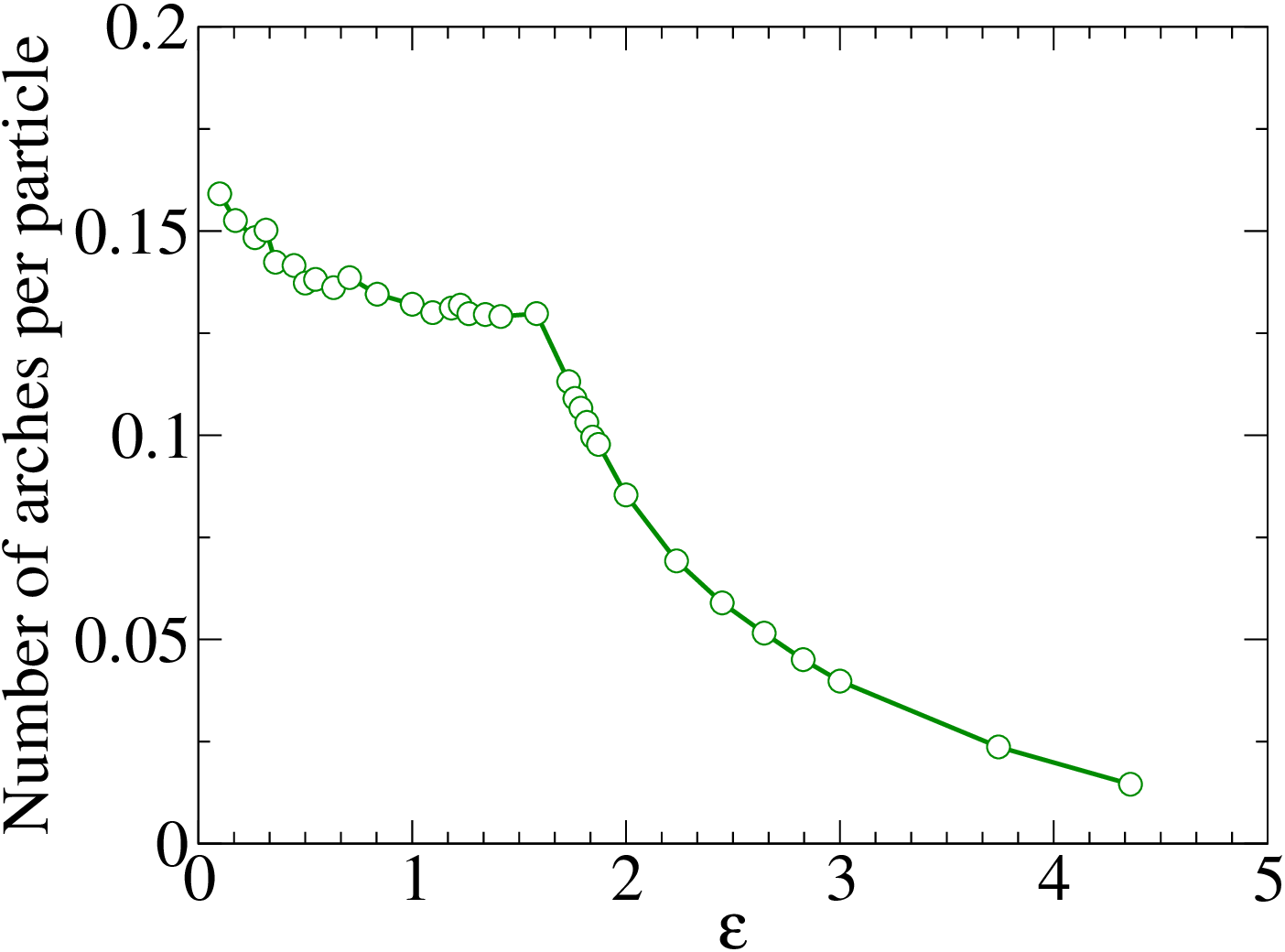}
\caption{\label{fig6} (Color on-line) Steady state number of arches for spheres as a function of the tapping intensity $\varepsilon$ for model A.}
\end{figure}

\begin{figure}
\includegraphics[width=0.45\textwidth]{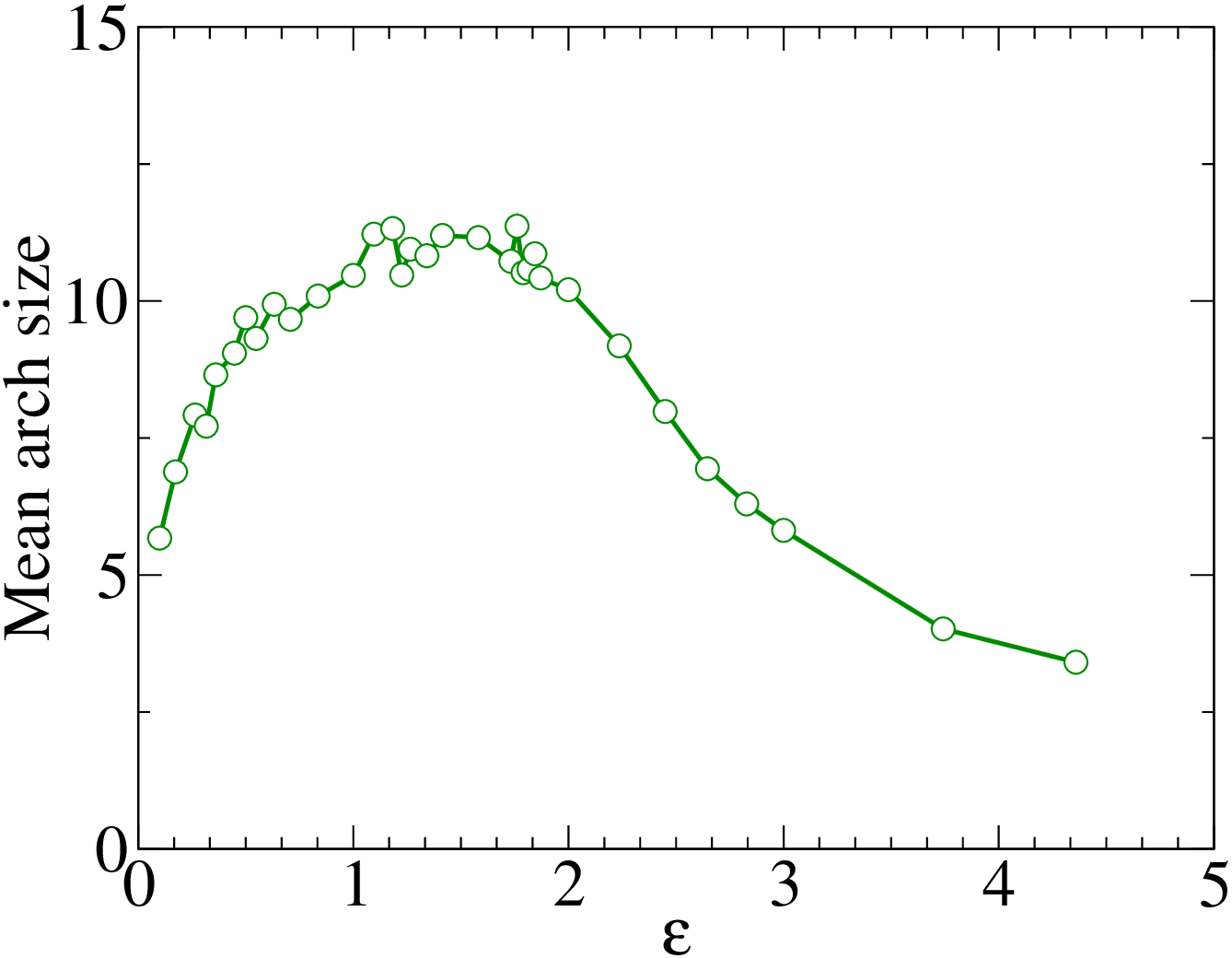}
\caption{\label{fig7} (Color on-line) Steady state mean arch size for spheres as a function of the tapping intensity $\varepsilon$ for model A.}
\end{figure}

Mehta and Barker \cite{Barker1,Mehta1} have suggested that arches are the main responsible for void trapping in shaken granular assemblies and that a proper model of the way arches are created and destroyed by tapping should lead to correct predictions of the behavior of $\phi$ versus $\varepsilon$. Arches are multiparticle structures where all particles are mutually stable, i.e. fixing the positions of all other particles in the assembly the removal of any particle in the arch leads to the collapse of the other particles in it. For an arch to be formed, it is necessary (although not sufficient) that two or more falling particles be in contact at the time they reach equilibrium in order to create mutually stabilizing structures. In FIGs. \ref{fig4}-\ref{fig7} we present results on the number and size of the arches found in our simulations for disks and spheres respectively. The mean arch size is calculated as the normalized second moment of the arch size distribution $\sum_{s=2}{s^2n(s)}/\sum_{s=2}{sn(s)}$. Here, $n(s)$ is the number of arches of $s$ particles. The sums run from the smallest possible arch size, i.e., $s=2$. We only consider models A, B and D since arches can be identified only in these models where contacts are effectively made between the particles and the history of the deposition is available. 

Details on the algorithms used to identify arches can be found in previous works \cite{Pugnaloni1,Arevalo1,Pugnaloni2,Pugnaloni3}. Briefly, we need first to identify the supporting grains of each particle in the packing. In 2D, there are two disks that support any given grain; in 3D, three contacting spheres are needed to support a given particle. A set of grains in contact with a given particle are able to provide support if the segment (in 2D) or the triangle (in 3D) defined by the contact points lies below the center of mass of the particle. Of course, some of these supporting contacts may be provided by the walls of the container. Then, we find all \textit{mutually stable particles}. Two grains A and B are mutually stable if A supports B and B supports A. Arches are defined as sets of particles connected through \textit{mutually stabilizing contacts}. The fact that the supporting particles of each grain have to be known implies that contacts, and the chronological order in which they happen, have to be clearly defined in the model. Unfortunately this information is not available in the simplest model C.

As we can see in FIGS. \ref{fig4} and \ref{fig5} the realistic model D presents a clear connection between arching and packing fraction. Both, the number of arches and the mean arch size, present a maximum at the tapping intensity where packing fraction is lower. Therefore, in our granular dynamics, an increase(decrease) in the steady state density is always associated with a decrease(increase) in the number and size of the arches; which is in accord with intuition.

The behavior of the pseudodynamic model (model B) is somewhat intriguing. While there is a maximum in the number (FIG. \ref{fig4}) and size (FIG. \ref{fig5}) of the arches, this is not coincident with the position of the minimum packing fraction. We find that in the range $0.4<\varepsilon<0.7$ the packing fraction falls even though the number of arches stays constant and the mean size of the arches falls. We presume that in this range the geometry of the arches plays an important roll. Arches of the same size in terms of number of particles may trap voids of different sizes depending on the particular arrangement. A detailed study of the geometrical changes undergone by arches in this regime will be presented elsewhere. Outside this peculiar regime, as in model D, we see that an increase(decrease) in $\phi$ is always associated with a decrease(increase) in the number and size of the arches. This time the changes in the number and size of arches are more pronounced than in model D, in correspondence with the steeper changes in $\phi$ observed in the pseudodynamics (see FIG. \ref{fig2}).

Finally, in FIGS. \ref{fig6} and \ref{fig7} the number and size of arches of the hybrid model (model A) displays an interesting behavior. In the range of $\varepsilon$ in which $\phi$ decreases the number of arches decreases very smoothly while the mean size of the arches grows significantly. Let us point out that the mean arch size for the 2D models are always within two and three (see FIG. \ref{fig5}) whereas for the 3D model the mean arch size grows from five (at the lowest $\varepsilon$ studied) up to twelve (when the minimum steady state density is achieved). Therefore, we argue that the well known decrease in the steady state packing fraction of vibrated granular assemblies is governed by a marked increase in the size of the arches with a mild reduction in the number of these structures. We have seen that the number of particles involved in arches stays constant along the initial decrease of $\phi$; since small arches are replaced by larger arches, the total number of these structures must fall slightly. Beyond the value of $\varepsilon$ at which the minimum $\phi$ is found, both the size and number of arches fall which leads to the final increase in the steady state packing fraction.

It is particularly simple to explain the high intensity tapping regime observed in our simulations by considering arching. When $\varepsilon$ is increased considerably, every tap expands the assembly in such a way that particles get well apart from each other. During deposition, particles will reach the free surface of the bed almost sequentially (one at a time) reducing the chances of mutual stabilization. Therefore, arches are less probable to form as $\varepsilon$ increases and so $\phi$ must grow since less voids get trapped. Indeed, we see that the number and size of arches decrease at large $\varepsilon$ for increasing tapping intensities (see FIGS. \ref{fig4} and \ref{fig5} for 2D and FIGS. \ref{fig6} and \ref{fig7} for 3D). Eventually, for very large $\varepsilon$ no arches are formed after each tap and $\phi$ reaches a limiting value. It is worth noting that in model D arches are never fully removed by an increase in $\varepsilon$. Since particles colliding with the free surface of the deposit induce others to bounce. In this way, rearrangements that lead to mutual stabilization and arching are always promoted.

The above description based on arching should, if one seeks to develop a consistent model, be able to explain the decrease of $\phi$ at low $\varepsilon$. This should lead to a justification for the existence of a minimum within the framework of the arching model. How could an increase in $\varepsilon$ induce the creation of more arches and/or larger arches that lead to lower $\phi$ in the low tapping intensity regime?  At low $\varepsilon$ the free volume injection due to a tap creates very narrow gaps between particles. For a given arch to grow by the insertion of a new particle, it is necessary to create a gap between two particles in the existing arch where the new particle can fit in. This explains why increasing $\varepsilon$ will initially promote the formation of larger arches. However, as we mentioned above, if taps are too strong particles will eventually deposit so separated from one another that they will not be able to form mutually stable structures.

\section{Conclusions}
In summary, we have shown that four independent model granular beds present a nonmonotonic dependence of the steady state packing fraction as a function of the tapping intensity. This suggests that the phenomenon exists no matter the details considered in the particle-particle interactions. A tentative explanation based on arching seems to be consistent with the size and number of arches detected in the simulated packings. 

We believe that the nonmonotonic behavior will be encountered in real experiments if tapping intensities can be raised beyond the values explored so far. To our knowledge, experiments with glass beads have surveyed up to $\Gamma \approx 10$. This involve expansion of the order of 1\% (i.e., $\varepsilon \approx 0.1$). It may be necessary to go well above these values (up to $\varepsilon \approx 2.0$, i.e., an expansion factor $A=5.0$) in 3D packings to observe the increase in packing fraction. Interestingly, in 2D packings, the minimum in $\phi$ may be encountered at $\varepsilon \approx 0.5$ (i.e., an expansion by a factor 1.25) which is much easier to achieve in a experimental set up. Since the high intensity tapping regime (where the increase in $\phi$ is observed) is achieved thanks to the particles settling separately without meeting each other in flight, one needs to achieve expansions large enough so that each grain will find its stable position without interacting with any other moving grain. In 2D, the number of neighbors is much smaller than in 3D. This suggest that relatively modest expansion will lead to fairly sequential deposition in 2D as compared with 3D. Hence, lower tapping intensities should be needed in 2D to find the minimum in the $\phi$-$\varepsilon$ curves as shown in the simulation results presented here.

The nonmonotonic behavior seems more evident if particles have a very low restitution coefficient according to the results of models A and B. We mention here a suggestive evidence of nonmonotonic behavior found in the literature. Experiments on 2D packings \cite{Blumenfeld1} deposited by means of a conveyor belt display an increasing packing fraction with decreasing density prior to deposition. The density of the free falling grains prior to deposition can be interpreted as the density of our simulated models A, B and C after the expansion that mimics the tapping. Therefore, the higher the tapping intensity, the lower the density prior to deposition, which leads to higher packing fractions. These experiments carried out in a horizontal setup, where particles move very slowly, effectively draw the restitution coefficient close to zero.

Finally, we point out that the simplest model discussed (model C) is able to show the same general trends displayed by the other models. This suggests that in order to explore other type of systems, such as non spherical grains, this model may suffice if only the relative changes in packing fraction is needed in a qualitative fashion. A recent study made on pentagons \cite{Vidales1} with the pseudodynamic model has shown that the $\phi$ vs. $\varepsilon$ curve presents a monotonic increase rather than a minimum. It would by interesting to see whether grains with other shapes present new features, and this could be achieved with the simple MC compaction approach. Moreover, it remains to be explored if much simpler models of granular compaction \cite{Nicodemi1,Prados1} do show a nonmonotonic reversible branch.

\begin{acknowledgments}

We are indebted to Gary C. Barker for his useful comments and for providing the source code of his hybrid deposition algorithm. The simulations involving the granular dynamics model where carried out with the algorithm written by Roberto Ar\'{e}valo. LAP, MC and FV acknowledge financial support from CONICET (Argentina). MM thanks CONICET for a scholarship.

\end{acknowledgments}

\end{document}